%% file: main.tex
\documentclass[sigconf]{acmart}

\AtBeginDocument{%
  }

\copyrightyear{2026}
\acmYear{2026}
\setcopyright{cc}
\setcctype{by}
\acmConference[WWW '26]{Proceedings of the ACM Web Conference 2026}{April 13--17, 2026}{Dubai, United Arab Emirates}
\acmBooktitle{Proceedings of the ACM Web Conference 2026 (WWW '26), April 13--17, 2026, Dubai, United Arab Emirates}
\acmPrice{}
\acmDOI{10.1145/3774904.3792810}
\acmISBN{979-8-4007-2307-0/2026/04}

\usepackage{enumitem,amsmath}
\usepackage{multirow}
\usepackage{subcaption}
\usepackage{makecell}
\usepackage{xcolor}
\usepackage{booktabs} 
\usepackage{graphicx}
\usepackage{algorithm}
\usepackage{algorithmic}

\newcommand{\spool}[1]{\mathrm{\text{Sum}}(#1)}

\begin{document}

\title{GRank: Towards Target-Aware and Streamlined Industrial Retrieval with a Generate-Rank Framework}

\author{Yijia Sun}
\authornote{These authors contributed equally to this work.} 
\email{sunyijia@kuaishou.com}
\affiliation{
\institution{Kuaishou Technology}
\city{Beijing}
\country{China}
}

\author{Shanshan Huang}  
\authornotemark[1]
\email{huangshanshan@kuaishou.com}
\affiliation{
\institution{Kuaishou Technology}
\city{Beijing}
\country{China}
}

\author{Zhiyuan Guan}
\authornote{This author conducted the research while at Kuaishou Technology. He is now with Shanghai Jiaotong University.}
\email{guanzhiyuan03@kuaishou.com}
\affiliation{
\institution{Kuaishou Technology}
\city{Beijing}
\country{China}
}

\author{Qiang Luo}
\authornote{Corresponding authors.} 
\email{luoqiang@kuaishou.com}
\affiliation{
\institution{Kuaishou Technology}
\city{Beijing}
\country{China}
}

\author{Ruiming Tang}
\authornotemark[3]
\email{tangruiming@kuaishou.com}
\affiliation{
\institution{Kuaishou Technology}
\city{Beijing}
\country{China}
}

\author{Kun Gai}
\email{gai.kun@qq.com}
\affiliation{
\institution{Unaffiliated}
\city{Beijing}
\country{China}
}

\author{Guorui Zhou}
\authornotemark[3]
\email{zhouguorui@kuaishou.com}
\affiliation{
\institution{Kuaishou Technology}
\department{Algorithm}
\city{Beijing}
\country{China}
}

\renewcommand{\shortauthors}{Yijia Sun et al.}

\begin{abstract}
Industrial-scale recommender systems rely on a cascade pipeline in which the \emph{retrieval} stage must return a high-recall candidate set from \emph{billions} of items under tight latency. Existing solutions either (i) suffer from limited expressiveness in capturing fine-grained user-item interactions, as seen in decoupled dual-tower architectures that rely on separate encoders, or generative models that lack precise target-aware matching capabilities, or (ii) build \emph{structured indices} (tree, graph, quantization) whose item-centric topologies struggle to incorporate dynamic user preferences and incur prohibitive construction and maintenance costs.

We present \textsc{GRank}, a novel \emph{structured-index-free} retrieval paradigm that seamlessly unifies target-aware learning with user-centric retrieval. Our key innovations include: (1) A \emph{target-aware Generator} trained to perform personalized candidate generation via GPU-accelerated MIPS, eliminating semantic drift and maintenance costs of structured indexing; (2) A \emph{lightweight but powerful Ranker} that performs fine-grained, candidate-specific inference on small subsets; (3) An end-to-end multi-task learning framework that ensures semantic consistency between generation and ranking objectives.

Extensive experiments on two public benchmarks and a billion-item production corpus demonstrate that GRank improves Recall@500 by over 30\% and 1.7× the P99 QPS of state-of-the-art tree- and graph-based retrievers.

\noindent\textsc{GRank} has been \textbf{fully deployed in production} in our recommendation platform since Q2 2025, serving 400 million active users with 99.95\% service availability. Online A/B tests confirm significant improvements in core engagement metrics, with Total App Usage Time increasing by 0.160\% in the main app and 0.165\% in the Lite version.
\end{abstract}



\maketitle

\input{intro_v2}
\input{rw_new}

\input{methods}

\input{experiments}
\input{Conclusion_short}

\bibliographystyle{ACM-Reference-Format}
\bibliography{sample-base}

\appendix
\input{appendix_mathmatic}

\input{appendix_hp_tables} 
\input{appendix_training_details}

\input{appendix_robust}

\end{document}

%% file: intro_v2.tex
\section{Introduction}
Recommender systems serve as the bedrock of modern information discovery, typically orchestrating a multi-stage cascade (Retrieval $\rightarrow$ Pre-ranking $\rightarrow$ Ranking $\rightarrow$ Re-ranking) to filter billons of items. Within this pipeline, the \textbf{retrieval stage} acts as the initial funnel. Its primary mandate is to efficiently identify a coarse candidate set containing the user's potential interests. The quality of this set dictates the upper bound of the entire system's performance.

To understand the evolution and challenges of retrieval models, it is essential to distinguish between two fundamental paradigms regarding how user preferences are modeled:

\textbf{1. Target-Agnostic Retrieval:} The user is encoded into a fixed representation vector $\mathbf{u}$ (e.g., via Dual-Tower models \cite{dssm,youtube_dnn,mind} or sequential transformers \cite{sasrec,bert4rec}), which is independent of any specific candidate item $\mathbf{v}$. Relevance is computed via a lightweight metric, typically the inner product $s(\mathbf{u}, \mathbf{v}) = \mathbf{u}^\top \mathbf{v}$. 

\textbf{2. Target-Aware Retrieval:} The user representation is dynamically refined or computed \textit{conditioned} on the specific target candidate $\mathbf{v}$, formulated as $s(\mathbf{u}, \mathbf{v}) = f_{\theta}(\mathbf{u}, \mathbf{v})$.

\textbf{Why is this necessary?} User interests are often polysemous and context-dependent. A user's history may contain diverse behaviors (e.g., clicking on both "gaming laptops" and "office chairs"). In a target-agnostic model, these interests are compressed into a single mean vector, often diluting specific intents. In contrast, a target-aware model can perform \textbf{candidate-specific refinement}: when scoring a "mouse," the model activates the "laptop" history; when scoring a "desk," it attends to the "chair" history. This ability to discern subtle user intents from noisy history is critical for high-precision retrieval. 

While target-aware models theoretically offer superior precision, their deployment is hindered by a prohibitive computational cost: executing a complex interaction function $f_{\theta}(\mathbf{u}, \mathbf{v})$ for every item in a billion-scale corpus is infeasible. 

To balance efficiency and precision, existing retrieval paradigms generally fall into two categories, each facing fundamental limitations:

\textbf{Paradigm 1: The Standard of Target-Agnostic Retrieval.}
Currently, the vast majority of industrial systems employ target-agnostic architectures, such as Dual-Encoders \cite{dssm,youtube_dnn,mind} or sequential transformers \cite{sasrec,bert4rec}. These models decouple user and item encoding, enabling highly efficient retrieval via Maximum Inner-Product Search (MIPS)

\textbf{Limitation:} The efficiency comes at a cost. By compressing a user's complex, multi-faceted interests into a single fixed vector $\mathbf{u}$ \textit{before} seeing any candidate, these models suffer from the "early fusion bottleneck." They preclude any candidate-specific refinement, often failing to retrieve relevant items that lie in the "long tail" of user interests or requiring an excessively large candidate set to maintain recall.

\textbf{Paradigm 2: The Complexity of Structured Indices.}
To bridge the gap between efficiency and target-aware precision, a separate line of research (e.g., TDM \cite{tdm}, JTM \cite{jtm}) incorporates target-aware scoring into \textbf{Structured Indices} (Trees or Graphs). Instead of scanning the full corpus, these methods organize items into a hierarchical structure and formulate retrieval as a beam search—traversing from coarse clusters to fine-grained items.

\textbf{Limitation:} While this enables target-aware pruning, it introduces severe engineering and methodological flaws:
\begin{itemize}[leftmargin=*,topsep=0pt,noitemsep]
    \item \textbf{Structural Rigidity:} These indices rely on \textit{item-centric} clustering (e.g., grouping items by category). This static structure often misaligns with fluid, dynamic user intents. If a user's intent does not match the pre-defined index path, the retrieval fails irrecoverably.
    \item \textbf{Engineering Overhead:} \
    textcolor{blue}{Maintaining tree/graph indices is operationally expensive. They suffer from structural imbalance leading to unpredictable latency (poor P99), and the complex offline construction process creates a bottleneck that hinders rapid model updates.}
\end{itemize}

\input{Grank_arch}

\vspace{0.5em}

\noindent\textbf{The Root Cause and Our Insight.} The identified limitations expose a fundamental deadlock in retrieval design: \textit{Efficiency} currently demands target-agnostic MIPS, while \textit{Precision} necessitates rigid structured indices. We argue this stems from a monolithic view that conflates two distinct objectives: \textbf{Global Pruning (Recall):} Efficiently narrowing a billion-scale corpus to a coarse candidate set. \textbf{Local Distinction (Precision):} Fine-grained intent disambiguation via user-item interaction. Existing structured indices attempt to force-fit both objectives into a single search path, incurring immense engineering overhead. \textbf{GRank} breaks this deadlock by decoupling these goals into a dedicated \textit{Generate-then-Rank} framework.

\vspace{0.5em}
\noindent\textbf{Our Solution: The GRank Framework.}
We propose \textsc{GRank}, a novel \textit{Generate-then-Rank} paradigm that decouples retrieval into two specialized, learnable stages, thereby eliminating the reliance on structured indices.
\begin{itemize}[leftmargin=*,topsep=0pt]
    \item \textbf{Stage 1 (Generator): Implicitly Target-Aware Pruning.} Instead of a static dual-tower, we design a generator trained with \textit{target-aware supervision}. Through a novel training paradigm, the generator absorbs the ability to anticipate candidate-specific preferences into the user representation. At inference, it retains the efficiency of standard MIPS to produce a compact candidate set, effectively bridging the gap between agnostic architecture and aware capability.
    \item \textbf{Stage 2 (Ranker): Explicitly Target-Aware Scoring.} A lightweight cross-attention network performs fine-grained interaction modeling on the small subset from Stage 1. This stage handles the heavy lifting of intent disambiguation without the latency burden of full-corpus scanning.
\end{itemize}
By shifting the complexity of target-aware modeling from \textit{inference-time indexing} to \textit{training-time learning}, \textsc{GRank} breaks the efficiency-precision dichotomy.

\vspace{0.5em}
\noindent\textbf{Key Contributions:}
\begin{itemize}[leftmargin=*,topsep=0pt]
    \item \textbf{Framework:} We propose a universal two-stage retrieval architecture that seamlessly unifies efficient generation and precise ranking in an end-to-end manner.
    \item \textbf{Methodology:} We introduce a novel training strategy that distills target-aware signals into a MIPS-compatible generator, achieving precision without indexing overhead.
    \item \textbf{Validation:} Extensive experiments show over 30\% improvement in Recall@500 and 1.7$\times$ QPS gain.
\end{itemize}
\vspace{0.5em}
\textsc{GRank} has been deployed in production since Q2 2025, serving 50 billion daily requests with 99.95\% availability. Online A/B tests confirm significant gains: \textbf{Total App Usage Time} increased by 0.160\% in the main app and 0.165\% in the Lite version.

%% file: Grank_arch.tex
\begin{figure*} %
    \centering
    \includegraphics[width=0.9\textwidth]{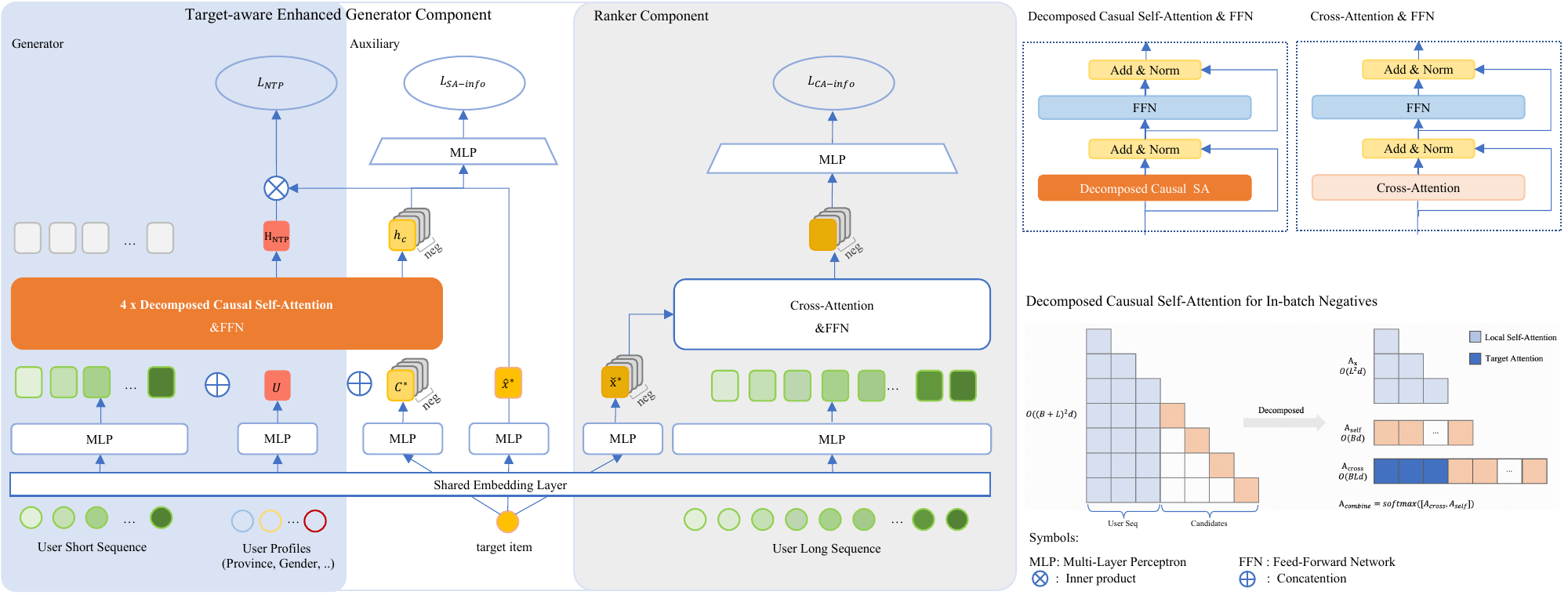} 
    \caption{GRank Framework Architecture. 
The framework comprises two tightly-coupled modules optimized through three complementary losses: (1) \textbf{Target-Aware Enhanced Generator} integrating \textbf{Generator} (optimized by $\mathcal{L}_{\text{NTP}}$) for short-term preference modeling and \textbf{Auxiliary} (optimized by $\mathcal{L}_{\text{SA-info}}$) for target injection; (2) \textbf{Ranker} (optimized by $\mathcal{L}_{\text{CA-info}}$) employing efficient cross-attention for fast inference. The Auxiliary module injects target embedding $\mathbf{C}^*$ and in-batch negatives $\mathbf{C_j}$ during training only. The Generator utilizes our novel \textbf{decomposed causal self-attention} (detailed in bottom-right) to efficiently process user sequences alongside candidate items while maintaining mathematical equivalence. All components are jointly optimized through multi-task learning.}
    \label{fig:model_arch}
\end{figure*}

%% file: rw_new.tex
\section{Related Work}
\subsection{Efficient yet Target-Agnostic Retrieval Paradigms}
\paragraph{Dual-Tower Architectures:}
Early retrieval systems predominantly employed \textit{dual-tower architectures} (e.g., DSSM \cite{dssm}, YouTube Recommendations \cite{youtube_dnn}), which independently encode users and items into embeddings $\mathbf{u} \in \mathbb{R}^d$ and $\mathbf{v} \in \mathbb{R}^d$, then measure relevance via a simple inner product enabling MIPS on GPUs \cite{faiss, 2019ann, 2024linr} or specialized ANN libraries \cite{scann} with logarithmic query time. However, this design is inherently \emph{target-agnostic}: the same user representation $\mathbf{u}$ is reused for all items, preventing candidate-specific refinement and limiting expressiveness for complex user intents \cite{covington2016deep}.

\paragraph{Generative Retrieval:}
Recent work reformulates retrieval as a \textit{generative sequence modeling} problem. This paradigm shift began with models like SASRec \cite{sasrec}, which employed unidirectional self-attention to model user sequences for next-item prediction.
BERT4Rec \cite{bert4rec} extended this by introducing bidirectional context encoding, while subsequent works like GPTRec \cite{gptrec} fully embraced an autoregressive framework. 
Further innovations such as Kuaiformer \cite{kuaiformer} and MPFormer \cite{sun2025mpformer} introduce multi‑interest tokens and adaptive sequence compression. Despite their sophistication, these methods remain fundamentally \emph{target‑agnostic}: the user representation is still computed independently of the candidate item (i.e., $\text{score}(\mathbf{u}, \mathbf{v}) = \phi(\text{Enc}_{\text{user}}(S), \mathbf{v})$), lacking explicit interaction mechanisms $g(\mathbf{u}, \mathbf{v})$ to focus history on specific candidates \cite{zhou2018deep, zhou2025survey}.

\subsection{Target-Aware Retrieval via Structured Indexing}
To address the target awareness problem, \textit{index-based retrieval} methods incorporating explicit interaction modeling have emerged. Key approaches include Tree-Based Indexing(such as TDM~\cite{tdm}, JTM\cite{jtm}, BSAT\cite{bsat}, MISS\cite{miss2025}), Graph-Based Indexing(NANN \cite{nann}, DR\cite{dr2021}), and Quantization-Based Indexing(StreamingVQ \cite{streamingvq}).

\subsubsection*{Tree-Based Indexing} 
Tree-based Deep Models propose tree structures to hierarchically search candidates. \textbf{TDM}~\cite{tdm} pioneers tree-based retrieval, reducing complexity to $O(\log N)$. However, its tree hierarchy $\mathcal{T}$ is typically \textit{decoupled} from the model objective $\mathcal{L}_{\text{model}}$. \textbf{JTM} \cite{jtm} alleviates the gap by co-training $\mathcal{T}$ and $\theta$ with heuristic clustering. However, the clustering itself is a limitation: user interest $\mathcal{D}_u$ can hardly perfectly aligns with item clusters $\mathcal{C}_i$, and the $O(N \log N)$ index rebuild is prohibitive for billion-scale item corpora $\mathcal{C}$ \cite{zhu2018learning}.

\subsubsection*{Graph-Based Indexing} 
\textbf{NANN}~\cite{nann} resorts to HNSW~\cite{hnsw} for candidate searching with model-generated edge weights. However, the construction of $\mathcal{G}$ often relies on pre-defined metrics (e.g. cosine similarity), which impose strong geometric assumptions on the item embedding space.  This deviation is exacerbated by non-linear encoders, causing retrieval paths to diverge from ranking objectives \cite{johnson2019billion}.

\subsubsection*{Quantization-Based Indexing}
\textbf{StreamingVQ} \cite{streamingvq} employs quantization clustering to compress high-dimensional embeddings into short codes using a codebook $\mathcal{B} = \{\mathbf{c}_1, ..., \mathbf{c}_K\}$. This enables sub-linear retrieval via table lookups. However, quantization indices typically assume Euclidean space ($\text{sim}(\mathbf{v}, \mathbf{c}_k) = -\|\mathbf{v} - \mathbf{c}_k\|^2_2$). Complex encoders (e.g., those using attention) break this assumption, \textit{exacerbating} the mismatch between the quantization objective $\mathcal{L}_{\text{VQ}}$ and $\mathcal{L}_{\text{rank}}$, further reducing retrieval accuracy \cite{jegou2010product}. Additionally, training a well‑balanced codebook is non‑trivial: items often concentrate in a few clusters, diminishing the pruning benefit of table lookups and introducing extra engineering overhead.

\subsubsection*{Fundamental Limitation: Item-Centric expansion}
Beyond these specific technical challenges, all structured indexing methods share a core architectural limitation: \textit{item-centric candidate expansion}. The search path is determined by pre-computed item-item proximities, inherently excluding real-time, personalized user signals during traversal. This results in semantic deviation where retrieval paths diverge from actual user intent \cite{rendle2019difficulty,chiu2019learning}, as user-specific context cannot dynamically re-route the query.

These limitations motivate our work \textsc{GRank}, which seeks to achieve target-aware precision without relying on structured indices or incurring their associated overheads.

%% file: methods.tex
\section{Methodology}
This section details the \textbf{GRank} framework, structured as follows. We begin by formalizing the retrieval task in §\ref{subsec:problem_formulation}. We then present our offline training methodology in §\ref{subsec:train_phase}, which jointly optimizes the Generator and Ranker modules end-to-end. Finally, §\ref{subsec:serve_phase} describes the efficient two-stage online inference pipeline. Figures~\ref{fig:model_arch} and ~\ref{fig:model_serving} illustrate the complete training and serving architectures, respectively.

\subsection{Problem Formulation}
\label{subsec:problem_formulation}

Let $\mathcal S_u = \{(i_1, f_1), \ldots, (i_n, f_n)\}$ be a user's chronological interaction sequence with items $i_t \in \mathcal{I}$ and features $f_t \in \mathbb{R}^d$ (watch duration, engagement type, author embedding). Given a target item $i^*$, the retrieval task is to estimating the relevance score $\mathcal{E}(i^* \mid \mathcal S_u)$, which defines the engagement probability:
\begin{equation}
p(i^* \mid \mathcal S_u) \propto \exp(\mathcal{E}(i^* \mid \mathcal S_u)).
\end{equation}
The objective is to efficiently retrieve the top-$K$ items from corpus $\mathcal{I}$ that maximize this probability.

\subsection{Offline Training}
\label{subsec:train_phase}
GRank consists of two tightly-coupled, end-to-end trainable modules (Figure~\ref{fig:model_arch}):
(1) \textbf{Target-Aware Enhanced Generator}, integrating \textbf{Generator} for short-term preference modeling and \textbf{Auxiliary} for target injection to enhance both Generator and Ranker; (2) a lightweight \textbf{Ranker} that performs fine-grained scoring.

\subsubsection{Input Representation}
\label{subsubsec:input_rep}
All components share a common item embedding layer $\mathbf{E} \in \mathbb{R}^{|\mathcal{I}| \times d}$. A personalized query token $\mathbf{U}$ aggregates user demographics and sum-pooled embeddings from recent behavior sequences (click, long-view, etc.), providing a fixed-size user context:
\[
\mathbf{U} = \left[ \mathbf{u}; \ \spool{\mathbf{Seq}_{rs}}; \ \spool{\mathbf{Seq}_{click}}; \ \spool{\mathbf{Seq}_{long\_view}} \right],
\]
where \(\mathbf{u} \in \mathbb{R}^d\) encodes user attributes. Each \(\mathbf{Seq}_{\cdot}\) stacks embeddings of the $L_s$ most recent items of a behavior type. Sum pooling \(\spool{\cdot}\) aggregates each sequence into a fixed-size vector: \(\spool{\mathbf{Seq}_{\cdot}} = \sum_{i=1}^{L_{s}} \mathbf{e}_i\).

\begingroup
\subsubsection{Target-Aware Enhanced Generator}
\label{subsubsec:enhanced_generator}
The generator's core innovation is to \textit{learn target-aware discrimination during training} while maintaining \textit{target-agnostic efficiency during inference}. It comprises:

\begin{itemize}[leftmargin=*]
    \item \textbf{Generator:} A causal Transformer that models short-term user sequences.
    \item \textbf{Auxiliary Module (Training-Only):} Injects target-item and in-batch negative signals into the sequence, enabling the model to learn candidate-aware representations.
\end{itemize}

\paragraph{Input Encoding \& Model Structure}
Historical interactions \((i_t, f_t)\) are encoded as $\mathbf{x}_t \in \mathbb{R}^d$. The generator takes the recent history $\mathbf{x}_{L_l-L_s+1:L_l}$ and the personalized query token $\mathbf{U}$ as input. To inject target-awareness during training, the target item $i^*$ and in-batch negatives $\mathcal{B}$ are transformed via an auxiliary MLP: $\mathbf{C}^* = \text{MLP}_{\text{aux}}(\mathbf{e}^*), \ \mathbf{C}_j = \text{MLP}_{\text{aux}}(\mathbf{e}_j)$, and appended to form the initial sequence:
\begin{equation}
\mathbf{H}^{(0)} = [\mathbf{x}_{L_l-L_s+1:L_l};\, \mathbf{U};\, \mathbf{C}],
\end{equation}
where $\mathbf{C} = [\mathbf{C}^*, \mathbf{C}_1, \ldots, \mathbf{C}_{|\mathcal{B}|}]$. This construction provides target-specific signals and enables efficient InfoNCE loss computation. The subsequent section (§\ref{subsubsec:train_acc}) will detail our optimized attention mechanism that efficiently processes this extended sequence while maintaining mathematically equivalent.

The sequence $\mathbf{H}^{(0)}$ is then processed by a causal Transformer decoder, with the user representation $\mathbf{h}_u$ extracted from the output corresponding to $\mathbf{U}$.

\paragraph{Training Objectives}
The generator is optimized through two complementary losses:
\begin{itemize}[leftmargin=*,noitemsep]
    \item \textbf{Generator Loss ($\mathcal{L}_{\text{NTP}}$):} 
    The user representation $\mathbf{h}_u$ is normalized by  $\ell_2$-norm to $\mathbf{H}_{\text{NTP}}$, then contrasted with the target and in‑batch negatives $\hat{\mathbf{x}}_j = \text{MLP}_{\text{ntp}}(\mathbf{e}_j)$ using InfoNCE:
    \begin{equation}
        \mathcal{L}_{\text{NTP}} = -\log\frac{\exp(\langle \mathbf{H}_{\text{NTP}}, \hat{\mathbf{x}}^*\rangle / \tau)}{\sum_{j\in\mathcal{B}}\exp(\langle \mathbf{H}_{\text{NTP}}, \hat{\mathbf{x}}_j\rangle / \tau)},
    \end{equation}
    where $\tau$ is the softmax temperature. 
    \item \textbf{Auxiliary Score Loss ($\mathcal{L}_{\text{SA-info}}$):} The Transformer outputs for the target $\mathbf{h}_{C^*}$ and negatives $\mathbf{h}_{C_j}$ are scored as $s_{\text{sa}_j} = \text{MLP}_{\text{sa}}(\mathbf{h}_{C_j})$, followed by InfoNCE:
    \begin{equation}
        \mathcal{L}_{\text{SA-info}} = -\log\frac{\exp(s_{\text{sa}})}{\sum_{j\in\mathcal{B}}\exp(s_{\text{sa}_j})}.
    \end{equation}
\end{itemize}

\paragraph{Training-Serving Consistency}
The causal mask ensures that while historical tokens can attend to the target during training, $\mathbf{C}^*$ is masked out during inference, preventing information leakage and maintaining strict training-serving consistency.

\subsubsection{Optimized Training via Decomposed Causal Self-Attention}
\label{subsubsec:train_acc}

While the generator architecture provides target-aware modeling capabilities, the conventional approach of concatenating user sequences with candidate items for self-attention results in $\mathcal{O}((L_s+B)^2d)$ complexity, creating a computational bottleneck during training.

We address this through a mathematically equivalent decomposition that maintains causal integrity while enabling efficient parallel computation.

\textbf{Traditional Approach}
The baseline computes full self-attention on the concatenated sequence of user history and candidate items:
\begin{equation}
\mathbf{S} = [\mathbf{X}; \mathbf{C}], \quad \mathbf{A} = \text{softmax}\left(\frac{\mathbf{S}\mathbf{W}_Q\mathbf{W}_K^\top\mathbf{S}^\top}{\sqrt{d}} \odot \mathbf{M}\right).
\end{equation}

\textbf{Optimized Decomposition}
Our method decouples attention computation into three coordinated components:

\noindent \textbf{1. User Sequence Self-Attention:}
\begin{equation}
\mathbf{Q}_s = \mathbf{X}\mathbf{W}_Q, \quad \mathbf{K}_s = \mathbf{X}\mathbf{W}_K, \quad \mathbf{V}_s = \mathbf{X}\mathbf{W}_V
\end{equation}
\begin{equation}
\mathbf{A}_{ss} = \text{softmax}\left(\frac{\mathbf{Q}_s\mathbf{K}_s^\top}{\sqrt{d}} \odot \mathbf{M}_s\right), \quad \mathbf{O}_s = \mathbf{A}_s\mathbf{V}_s
\end{equation}

\noindent \textbf{2. Candidate Attention Decomposition:}
\begin{equation}
\mathbf{Q}_c = \mathbf{C}\mathbf{W}_Q, \quad \mathbf{K}_c = \mathbf{C}\mathbf{W}_K, \quad \mathbf{V}_c = \mathbf{C}\mathbf{W}_V
\end{equation}

Cross-Scores (Candidates → User Sequence):
\begin{equation}
\mathbf{A}_{\text{cs}} = \frac{\mathbf{Q}_c\mathbf{K}_X^\top}{\sqrt{d}} \in \mathbb{R}^{B \times L_s}
\end{equation}

Candidate Self-Correlation Scores:
\begin{equation}
\mathbf{A}_{\text{cc}} = \frac{\sum_{i=1}^d \mathbf{Q}_{c,:,i} \odot \mathbf{K}_{c,:,i}}{\sqrt{d}} \in \mathbb{R}^{B \times 1}
\end{equation}

\noindent \textbf{3. Attention Combination and Output:}
\begin{equation}
\mathbf{A}_{\text{combined}} = \text{softmax}\left([\mathbf{A}_{\text{cs}}, \mathbf{A}_{\text{cc}}]\right), \quad \mathbf{V}_{\text{combined}} = [\mathbf{V}_v; \mathbf{V}_c]
\end{equation}
\begin{equation}
\mathbf{O}_c = \mathbf{A}_{\text{combined}}\mathbf{V}_{\text{combined}}, \quad \mathbf{O}_{\text{final}} = [\mathbf{O}_s; \mathbf{O}_c]
\end{equation}

\textbf{Complexity Analysis}
This decomposition reduces complexity from $\mathcal{O}((L_s+B)^2d)$ to $\mathcal{O}(L_s^2d + BL_sd + Bd)$, achieving 82\% FLOPs reduction under typical configurations ($B=300$, $L_s=64$, $d=128$) while maintaining mathematical equivalence with the traditional approach.
\endgroup

\textit{The mathematical equivalence proof and detailed algorithm analysis are provided in Appendix~\ref{app:attention_equivalence}.}

\subsubsection{Ranker Module}
The Ranker's role is to perform precise, candidate-specific scoring on the small candidate set from the Generator. Its design prioritizes effectiveness within a low compute budget.

\textbf{Encoding Long-Term Context and Candidates:} It processes an extended user history (up to 1000 items) through an MLP to capture long-term preferences $\mathbf{H}_u^{\text{long}} \in \mathbb{R}^{L_l \times d}$. Each candidate item is similarly transformed via a separate MLP to obtain an enriched query representation $\check{\mathbf{x}}^* \in \mathbb{R}^{d}$.

\textbf{Lightweight Cross-Attention:} For each candidate, a single-head cross-attention mechanism uses the candidate as query and $\mathbf{H}_u^{\text{long}}$ as keys/values, enabling fine-grained, target-aware interaction. The output is projected to a relevance score $s_{\text{ca}}$.

\textbf{Optimization:} The Ranker is trained with an InfoNCE loss $\mathcal{L}_{\text{CA-info}}$ using in-batch negatives. Its simplicity keeps inference latency minimal, as it only runs on hundreds (not millions) of items:
\begin{equation}
\mathcal L_{\text{CA-info}} = -\log\frac{\exp\!\bigl(s_{\text{ca}})}
{\sum_{j\in\mathcal B}\exp\!\bigl(s_{\text{ca}_j})}.
\end{equation}

\subsubsection{Multi-Task Training}
The Generator (with Auxiliary) and Ranker are trained jointly end-to-end:
\begin{equation}
\mathcal L_{\text{total}} = \lambda_0\mathcal L_{\text{NTP}} + \lambda_1\mathcal L_{\text{SA-info}} + \lambda_2\mathcal L_{\text{CA-info}},
\end{equation}
where weights $\lambda_i$ are tuned on a validation set. This ensures both stages are co-adapted: the Generator learns to produce candidates that the Ranker can accurately discriminate.

\subsection{Online Inference}
\label{subsec:serve_phase}
Online inference adopts a two-stage latency-sensitive cascade as shown in Figure ~\ref{fig:model_serving}.

\paragraph{Stage-1: NTP Generator.}
The vector $\mathbf H_{NTP}$ is used as a query for MIPS over a quantized embedding index, retrieving a top-$k_1=2{,}000$ candidate set $\mathcal C_1$ in $\mathcal O(\log |\mathcal I|)$ time.

\paragraph{Stage-2: Ranker.}
For each $i\in\mathcal C_1$, we compute the cross-attention score $s_{\text{CA}}$ and rerank the candidates to obtain the final top-$k_2=500$ items. This computation is purely feedforward and highly parallelizable, enabling high-throughput, low-latency inference when batched on GPU—making it well-suited for real-time serving (see §\ref{subsec:performance} for detailed benchmarks).

\input{fig_model_serving}

%% file: fig_model_serving.tex
\begin{figure}[thbp] %
    \centering
    \includegraphics[width=0.48\textwidth]{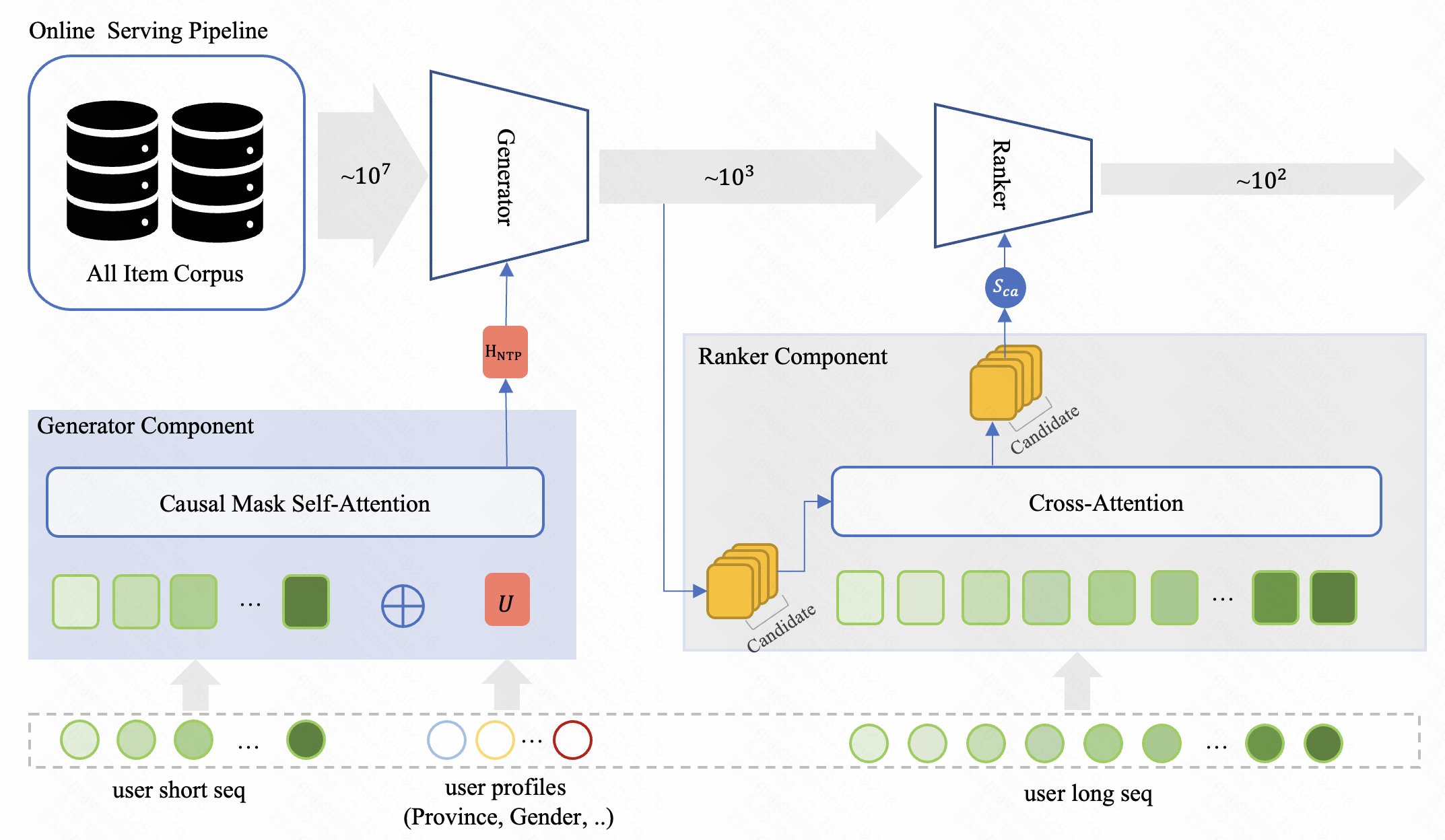} 
    \caption{GRank Serving Architecture. Stage-1 (Generator): Leverages user latent interests to generate a personalized candidate subset. Starge-2 (Ranker):Employs a target-aware cross-attention scorer for fine-grained relevance estimation. }
    \label{fig:model_serving}
\end{figure}

%% file: experiments.tex
\section{Experiments}

We conducted rigorous experiments to answer five research questions (RQs).

\begin{itemize}[leftmargin=*,nosep]
  \item \textbf{RQ1}: How effective is GRank compared with state-of-the-art models in the industry?
  \item \textbf{RQ2}: Does the GRank architecture meet industrial latency constraints?
  \item \textbf{RQ3}: What is the contribution of each core architectural component to the overall performance?
  \item \textbf{RQ4}: How do hyper-parameter selections enable the optimal trade-off between efficiency and performance in industrial deployment?
  \item \textbf{RQ5}: Does GRank bring significant online gains in our short-video service?
\end{itemize}

\input{table1_baseline_cmp}

\subsection{Experimental Setup}
\subsubsection{Datasets and Evaluation Protocol}
\label{subsubsec:datasets}

We evaluate GRank on one industrial and two public datasets (Table~\ref{tab:dataset_stats}): 
\textbf{MovieLens-20M} (movie recommendations), \textbf{Taobao UserBehavior} (e-commerce), and \textbf{Kuaishou Industrial} (short-video platform). All datasets are split chronologically: Kuaishou uses 6 days for training and the next day for testing, while the public datasets follow standard 90\%/10\% split.

\begin{table}[h]
\centering
\caption{Dataset statistics after preprocessing}
\label{tab:dataset_stats}
\begin{tabular}{lccc}
\toprule
 & \textbf{UserBehavior} & \textbf{MovieLens-20M} & \textbf{Kuaishou} \\
\midrule
Users & 964K & 138K & 108M \\
Items & 4.2M & 27K & 42M  \\
Interactions & 1.7M & 9.3M & 4B \\
Avg. Seq. Len. & 101 & 144 & 980 \\
\bottomrule
\end{tabular}
\vspace{-1em}
\end{table}

\subsubsection{Evaluation Metrics and Baselines}
We report \textbf{Recall@K} and \textbf{NDCG@K} (retrieval from full corpus), and \textbf{QPS} (maximum throughput under P99 latency $\leq$ 100ms). 

Baselines represent key retrieval paradigms: \textbf{DSSM} (dual-tower), \textbf{Kuaiformer} (generative), \textbf{TDM} (tree-based), \textbf{NANN} (graph-based), and \textbf{StreamingVQ} (quantization). All use identical training data, embeddings ($d=128$), and hardware configurations.

\subsubsection{Implementation Details}
\label{subsubsec:impl_details}
GRank uses embedding dimension $d=128$, 4-layer causal self-attention, 1-layer cross-attention, sequence lengths of 64 (short-term) and 1000 (long-term), candidate set size 2000. All baselines use identical embeddings and hardware for fair comparison. Comprehensive training, optimization, and deployment details are provided in Appendix~\ref{app:training_details}.

\subsection{Offline Evaluation (RQ1)}
\label{subsec:offline_results}
As summarized in Table~\ref{tab:baseline_cmp}, GRank achieves state-of-the-art performance across all datasets, demonstrating the effectiveness of its unified, index-free paradigm.

Our results demonstrate two key findings: First, GRank significantly outperforms target-agnostic models including generative (Kuaiformer) and dual-tower (DSSM) approaches, improving Recall@50 by \textbf{33.0\%} on the User Behavior dataset. All target-aware methods (TDM, NANN, StreamingVQ) consistently outperform those baselines on industrial data, confirming the necessity of target-aware modeling.

Second, GRank achieves superior performance even compared to sophisticated target-aware systems that rely on structured indices, with \textbf{32.8\%} higher Recall@500 than tree-based TDM on the Industry Dataset. The fact that GRank attains this superior accuracy without any structured index highlights the sufficiency of a well-trained generator combined with a lightweight ranker.

These consistent gains across diverse datasets confirm GRank's effectiveness in bridging the accuracy-efficiency gap in large-scale retrieval.

\input{fig_high_parameter_merge}

\subsection{Online Service Performance (RQ2)}
\label{subsec:performance}
This section evaluates whether GRank's two-stage architecture meets industrial latency constraints (RQ2). All experiments used production servers with identical GPU configurations, measuring QPS under the P99 latency constraint of $\leq 100$ ms. Reported latencies include the full pipeline from feature extraction to final scoring.

As shown in Table~\ref{tab:latency_throughput}, GRank achieves a QPS of 767, significantly outperforming NANN and TDM while maintaining superior recall. The suboptimal throughput of NANN and TDM can be attributed to path uncertainty, candidate set size fluctuations, and multi-round retrieval overhead. In contrast, GRank's efficiency stems from its two-stage design: the generator rapidly reduces candidate sets via efficient MIPS, avoiding complex tree/graph traversals, while the lightweight ranker applies precise target-aware scoring with minimal overhead. This approach avoids the latency variability of structured indices while delivering higher accuracy than target-agnostic models.
\input{table_latency_qps}

\subsection{Ablation Study on Architectural Components (RQ3)}
\label{subsec:ablation_arch}
To evaluate the contribution of each component within \textsc{GRank}, we design three strategic variants to address specific architectural questions:
\begin{itemize}[leftmargin=*, noitemsep, topsep=3pt]
    \item \textbf{Pure Generator}: \textit{How does the base generator perform as a standalone retrieval baseline?} This variant utilizes only the first-stage generator without any auxiliary supervision or ranking refinement, serving as the fundamental baseline.
    \item \textbf{Gen+Aux}: \textit{To what extent does auxiliary supervision enhance the generator's representation capability?} This variant excludes the cross-attention Rank Module. In the inference stage, it utilizes the output of the auxiliary module as a scoring mechanism for candidate refinement.
    \item \textbf{Gen+RankOnly}: \textit{Can cross-attention offer a superior efficiency-accuracy trade-off?} This variant augments the base generator with our proposed Rank Module but excludes the auxiliary supervision, directly evaluating the impact of the target-aware cross-attention mechanism.
\end{itemize}
\vspace{0.5em} %
Ultimately, the full \textsc{GRank} model demonstrates whether a hybrid architecture can synergize these components to achieve an optimal performance balance.

\input{ablation_attn_arch2}

\input{genrank_online_ab}

The ablation results conclusively demonstrate that our two-stage design with hybrid attention mechanisms achieves the best trade-off between retrieval quality and computational efficiency.
\begin{itemize}[leftmargin=*, topsep=3pt, itemsep=3pt]
    \item \textbf{Universal Offline-Enhancement Paradigm for Generator}. To further elucidate the performance gains of GRank, we dissect the behaviour of the Generator itself. Since the 2\,000 candidates produced in the first stage form the \emph{absolute} ceiling for any subsequent ranker, the jump of Generator Recall@2000 from 0.1512 to 0.3685 \emph{significantly contributes} to the final Recall@500 of 0.2346. 
    This observation reveals that injecting a target-aware ranking loss during training substantially strengthens the user representation of a dual-tower model, while leaving the tower architecture \emph{completely untouched} at inference—introducing \textbf{zero} additional latency. 
    Consequently, the ``\textbf{offline optimise, transparent deploy}'' paradigm acts as a drop-in, model-agnostic performance booster that is instantly applicable to \emph{any} representation-based or dual-tower retrieval system, delivering higher accuracy with no extra parameters, no extra engineering, and no extra cost at serving time.
    \item \textbf{Second-Stage Rescoring is Critical for Accuracy Gains}. The removal of rescoring drastically impairs performance (Recall@500: -50.7\%), highlighting its necessity. All rescoring variants deliver substantial metric gains over the pure generator, confirming the value of second-stage refinement, though at the cost of reduced QPS. The full GRank model achieves the most favorable accuracy-efficiency trade-off.

    \item \textbf{CA-SA Synergy Outperforms Isolated Pathways}.Our ablation shows that combining CA and SA outperforms either approach alone, with each mechanism offering distinct advantages:
    \begin{itemize}[leftmargin=*, nosep]
        \item \textbf{Generator with Self-Attention (Gen+Aux)}: This configuration achieves a Recall@500 of 0.1768, representing a significant improvement over the pure generator baseline. However, the quadratic computational complexity ($O(4*n^2)$) of 4 layers self-attention leads to substantial latency overhead.
        
        \item \textbf{Generator with Cross-Attention (Gen+RankOnly)}: This approach also demonstrates notable improvement in Recall@500 (0.1363) over the baseline. More importantly, CA's linear complexity ($O(n)$) enables efficient processing, particularly advantageous for long-sequence modeling tasks. Nevertheless, its accuracy remains inferior to the SA-enhanced variant. 
    
        \item \textbf{Hybrid CA-SA synergy optimizes trade-offs(Full GRank):} The complete GRank framework leverages both attention mechanisms: CA for efficient global modeling and SA for capturing fine-grained sequential dependencies during training and influence the update dynamics of shared sparse representations. Crucially, SA is excluded from inference, preserving the latency benefits of CA while maintaining superior accuracy.
    \end{itemize}
\end{itemize}

\subsection{Balancing Efficiency and Performance through Hyper-Parameter Selection (RQ4)}
\label{subsec:rq3_hp}

We systematically examine how hyper-parameters govern the trade-off between efficiency and performance in industrial deployment. Rather than merely reporting sensitivity, we identify optimal configurations that achieve the best practical balance for our production environment. Figure~\ref{fig:hp_merge} 
presents the hyper-parameter analysis, with detailed numerical data provided in  Appendix~\ref{app:hp_tables}  (Tables~\ref{tab:seq_length} --  ~\ref{tab:embed_dim}).

\subsubsection{Sequence Length: Long-Range Modeling with Linear Complexity}
\label{subsubsec:seq_len}
The analysis focuses on the CA module's sequence length. To maximize efficiency, scoring is performed exclusively by the CA module during inference. As shown in Figure~\ref{fig:hp_merge}.a, the performance improves consistently with increasing $L$: when $L=1,000$, Recall@500 reaches 0.2345, significantly surpassing the SA module's performance(0.1768) while maintaining substantially lower latency. This demonstrates that increasing sequence length effectively compensates for excluding the SA module at inference, achieving an optimal balance where enhanced effectiveness meets practical efficiency constraints.

\subsubsection{Candidate Size: ROI-Optimal Tuning}
\label{subsubsec:candidate_size}
The generator candidate size $k_1$ balances recall and throughput (Fig.~\ref{fig:hp_merge}.b). For our latency budget, $k_1=2,000$ is empirically optimal: beyond this, recall gains diminish sharply (only 4.27\% at $k_1=5,000$) while QPS drops 18\%, confirming its ROI-optimal status. This aligns with \textsc{GRank}'s two-stage design—the Generator establishes a high-recall ceiling, leaving fine-grained discrimination to the Ranker. Fig.~\ref{fig:hp_merge}.b provides a general performance–latency tradeoff curve; a practical tuning starting point is $k_1 = 2 \times n_\text{downstream}$, adjustable for recall or latency priorities.

\subsubsection{Embedding Dimension: Performance Saturation under Memory Constraints}
\label{subsubsec:top_embed}
We evaluate the selection of the top embedding dimension $d_{\text{top}}$ under production memory constraints. As shown in Figure~\ref{fig:hp_merge}.c, recall increases monotonically with increasing $d_{\text{top}}$ and saturates at 128, but drops significantly at 256 due to memory constraints that limit  candidate pool size. Meanwhile, $d_{\text{top}}=64$ maintains comparable performance while significantly improving system throughput. Therefore, $d_{\text{top}}=128$ is optimal for accuracy-critical scenarios, while $d_{\text{top}}=64$ provides an ideal alternative for throughput-sensitive deployments requiring efficiency trade-offs.

\subsection{Online Results (RQ5)}
\label{sec:online_results}
We conducted a one-week A/B test on Kuaishou's single-column recommendation platforms by replacing the original \textsc{Kuaiformer} recall with \textsc{GRank} while maintaining identical experimental conditions. The results in Table~\ref{tab:genrank_combined} demonstrate \textsc{GRank}'s effectiveness across both business and technical dimensions.

\noindent\textbf{Overall App Performance.}
\textsc{GRank} delivers consistent improvements in core user engagement metrics, with particular strength in video consumption and preference modeling. The framework effectively translates retrieval enhancements into measurable business value, with both main and lite versions showing positive trends across all key indicators.

\noindent\textbf{Recall Quality.}
\textsc{GRank} improves recall effectiveness, with higher show ratio, UV coverage, and average watch time per session, reflecting improved content relevance and increased user reach. These consistent improvements demonstrate \textsc{GRank}'s ability to enhance the entire recommendation ecosystem through superior retrieval.

The correlated improvements across both business metrics and technical indicators establish that \textsc{GRank}'s architectural innovations directly translate to enhanced user experience and platform value.

%% file: table1_baseline_cmp.tex
\begin{table*}[t]
\centering
\caption{Performance Comparison of SOTA Methods on Different Datasets. The best and second-best results highlighted in bold font and \underline{underlined}. \textit{Note: TDM was only evaluated on the Industry Dataset due to its deployment complexity, using our existing infrastructure.} }
\label{tab:baseline_cmp}
\begin{tabular}{lcccccccccc}
\toprule
\multirow{2}{*}{Method} & \multicolumn{2}{c}{User Behavior} & \multicolumn{2}{c}{MovieLens} & \multicolumn{2}{c}{Industry Datasets} \\
\cmidrule(lr){2-3} \cmidrule(lr){4-5} \cmidrule(lr){6-7}
& Recall@50 & NDCG@50  & Recall@50 & NDCG@50 & Recall@500 & NDCG@500\\
\midrule
DSSM  & 0.2711 & 0.1911 & 0.1940 & 0.0792 & 0.1068 & 0.0360 \\
Kuaiformer  & \underline{0.3270} & \underline{0.2347} & 0.2538 & 0.0906 & 0.1151 & 0.0386\\
TDM & - & - & - & - & \underline{0.1766} & 0.0535 \\
NANN & 0.3264 & 0.1765 & \underline{0.2633} & \underline{0.1017} & 0.1326 & 0.0466 \\
Streaming VQ & 0.3220 & 0.2262 & 0.2554 & 0.0907 & 0.1296 & \underline{0.0603} \\
\midrule
GRank  & \textbf{0.4348} & \textbf{0.2780}  & \textbf{0.3350} & \textbf{0.1250} & \textbf{0.2346} & \textbf{0.0775}     \\
Relative Gain & +33.0\% & +18.4\% & +27.2\% & +22.9\% & +32.8\% & +28.5\% \\

\bottomrule
\end{tabular}
\end{table*}

%% file: fig_high_parameter_merge.tex
\begin{figure*}[th]
\centering
\includegraphics[width=\linewidth]{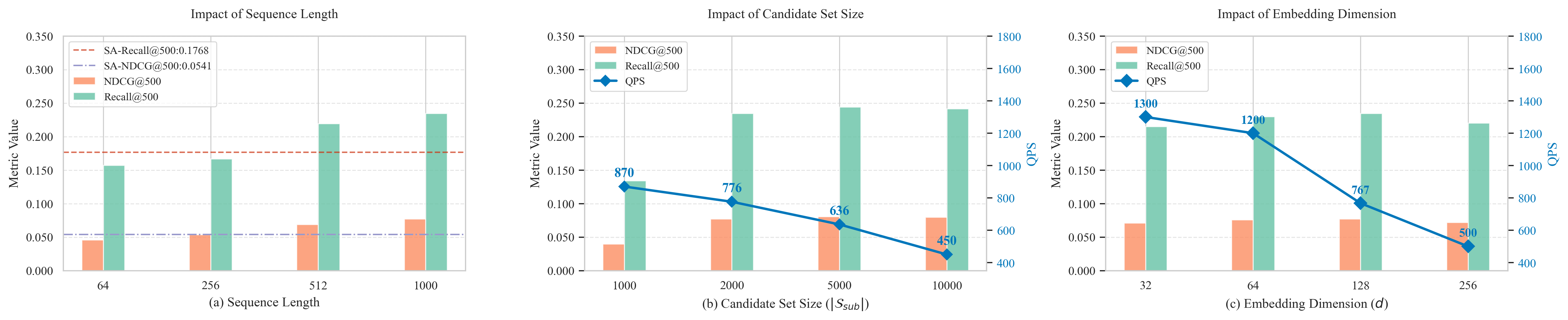}
  \caption{Hyper-parameter analysis on the efficiency-performance trade-off. 
  (\textbf{a}) Sequence length: CA Recall@500 increases monotonically with $L$, peaking when $L\!=\!1\,000$, significantly outperforming the SA module while maintaining lower latency.
  (\textbf{b}) Candidate size: diminishing returns begin at $k_1\!=\!2\,000$; recall improves by only $0.1$ pp at $k_1\!=\!5\,000$ while QPS drops $18\%$, making $2\,000$ the pragmatic optimum.
  (\textbf{c}) Top embedding dimension: recall saturates at $d_{\text{top}}\!=\!128$ and falls at $256$ owing to memory pressure; $d_{\text{top}}\!=\!64$ offers a high-ROI fallback, trading $3\%$ recall for $56\%$ QPS gain.}
\label{fig:hp_merge}
\end{figure*}

%% file: table_latency_qps.tex
\begin{table}[h]
\centering
\small
\caption{End-to-end latency, throughput and recall comparison. SLA threshold is 100ms. All latency values include full retrieval service pipeline (feature processing + core retrieval logic).}
\label{tab:latency_throughput}
\begin{tabular}{lcccccc}
\toprule
\textbf{Method} & \textbf{\makecell{Recall \\ @500} } & \textbf{\makecell{NDCG \\ @500} } & \textbf{QPS}  & \textbf{\makecell{P50 Lat. \\ (ms)}} & \textbf{\makecell{P99 Lat. \\ (ms)}} &  \\
\midrule
DSSM        & 0.1068 & 0.0360 & 1300 & 50  & 61  \\
Kuaiformer  & 0.1151 & 0.0386 & 772 & 59  & 87   \\
\midrule
TDM         & 0.1766 & 0.0535 & 273 & 55  & 96   \\
NANN        & 0.1326 & 0.0466 & 384 & 70  & 100  \\
StreamVQ    & 0.1296 & 0.0603 & 450  & 48  & 75  \\
\midrule
\textbf{GRank} & \textbf{0.2346} &  \textbf{0.0775} & \textbf{767} & \textbf{48} & \textbf{73} \\
\bottomrule
\end{tabular}
\end{table}

%% file: ablation_attn_arch2.tex
\begin{table}[htbp]
\centering
\caption{Ablation Results of Hybrid Attention Architecture}
\label{tab:ablation_results}
\begin{tabular}{lcccc}
\toprule
\textbf{Config} & \textbf{\makecell{Generator \\ Recall@2000}} & \textbf{\makecell{Recall \\ @500}} & \textbf{\makecell{NDCG \\ @500}} & \textbf{\makecell{QPS}} \\
\midrule
Pure Generator & 0.1512 & 0.1190 & 0.0405 & 1333  \\
Gen+Aux & 0.3427 & 0.1768 & 0.0541 & 340 \\
Gen+RankOnly & 0.1820 & 0.1363 & 0.0502 & 767 \\
\textbf{Full GRank}   & \textbf{0.3685} & \textbf{0.2346} & \textbf{0.0775}  & \textbf{767} \\
\bottomrule
\end{tabular}
\end{table}

%% file: genrank_online_ab.tex
\begin{table*}[ht]
\centering
\caption{Online A/B Test Results of GRank Across Different Scenarios}
\label{tab:genrank_combined}
\begin{tabular}{lccccccc}
\toprule
\multirow{2}{*}{Applications} & \multicolumn{4}{c}{Overall App Metrics} & \multicolumn{3}{c}{Recall Pathway Metrics} \\
\cmidrule(lr){2-5} \cmidrule(lr){6-8}
 & \makecell{Total App \\ Usage  Time} & \makecell{Usage Time \\ per User} & \makecell{Video \\ Watch Time} & \makecell{Effective \\ Interests} & \makecell{Show \\ Ratio} & \makecell{Avg. \\ Watch Time} & \makecell{UV \\ Coverage} \\
\midrule
Kuaishou Single Page  & +0.160\% & +0.139\% & +0.347\% & +0.527\% & +21.92\% & +16.26\% & +16.50\% \\
Kuaishou Lite Single Page  & +0.165\% & +0.118\% & +0.233\% & +0.384\% & +20.31\% & +11.69\% & +12.29\% \\
\bottomrule
\end{tabular}
\end{table*}

%% file: Conclusion_short.tex
\section{Conclusion}
In this paper, we presented \textbf{GRank}, a novel retrieval framework that fundamentally redefines the balance between efficiency and precision in large-scale recommendation. By decomposing the retrieval process into a decoupled \textbf{Generate-then-Rank} paradigm, \textsc{GRank} effectively bridges the gap between target-agnostic efficiency and target-aware expressiveness. Our approach demonstrates that high-precision, candidate-specific interaction modeling can be achieved without the rigid constraints and operational burdens of structured indices, such as trees or graphs.

Extensive experiments demonstrate over 30\% improvement in Recall, validating the framework’s effectiveness. GRank has been deployed in production since Q2 2025, now serving 400 million users and handling 50 billion daily requests, confirming its scalability and real-world impact. Ultimately, \textsc{GRank} provides a scalable and index-free alternative for modern retrieval systems, paving the way for more flexible and intelligent candidate generation in the industry.

%% file: appendix_mathmatic.tex
\section{Mathematical Equivalence of Decomposed Causal Self-Attention}
\label{app:attention_equivalence}

This appendix provides the formal mathematical foundation for the decomposed causal self-attention mechanism described in Section 3.2.3. Here we use consistent notation with the main text: $L$ denotes historical sequence length, $B$ denotes candidate batch size, and $d$ denotes hidden dimension.

The decomposition presented in Section 3.2.3 splits the input into historical ($\mathbf{X} \in \mathbb{R}^{L \times d}$) and candidate ($\mathbf{C} \in \mathbb{R}^{B \times d}$) components. In this appendix, we prove that this decomposition is mathematically equivalent to standard causal attention.

\subsection{Notation and Preliminaries}
Let $\mathbf{Q}$, $\mathbf{K}$, $\mathbf{V} \in \mathbb{R}^{L \times d}$ be the query, key, and value matrices, respectively, where $L$ is the sequence length and $d$ is the hidden dimension. The standard scaled dot-product attention is defined as:

\begin{equation}
\operatorname{Attention}(\mathbf{Q}, \mathbf{K}, \mathbf{V}) = \operatorname{softmax}\left(\frac{\mathbf{Q}\mathbf{K}^\top}{\sqrt{d}}\right) \mathbf{V}
\label{eq:standard_attention}
\end{equation}

For causal self-attention, we apply a causal mask $\mathbf{M} \in \{0,-\infty\}^{L \times L}$ where $M_{ij} = 0$ if $i \geq j$ and $M_{ij} = -\infty$ otherwise:

\begin{equation}
\operatorname{CausalAttention}(\mathbf{Q}, \mathbf{K}, \mathbf{V}) = \operatorname{softmax}\left(\frac{\mathbf{Q}\mathbf{K}^\top}{\sqrt{d}} + \mathbf{M}\right) \mathbf{V}
\label{eq:causal_attention}
\end{equation}

\subsection{Decomposition of Causal Self-Attention}
Our decomposed attention splits the sequence into two parts: historical items ($S$, length $L_s$) and candidate items ($C$, length $L_c$), such that $L = L_s + L_c$. Let:

\begin{align}
\mathbf{Q} &= \begin{bmatrix} \mathbf{Q}_s \\ \mathbf{Q}_c \end{bmatrix}, &
\mathbf{K} &= \begin{bmatrix} \mathbf{K}_s \\ \mathbf{K}_c \end{bmatrix}, &
\mathbf{V} &= \begin{bmatrix} \mathbf{V}_s \\ \mathbf{V}_c \end{bmatrix}
\end{align}

where $\mathbf{Q}_s, \mathbf{K}_s, \mathbf{V}_s \in \mathbb{R}^{L_s \times d}$ and $\mathbf{Q}_c, \mathbf{K}_c, \mathbf{V}_c \in \mathbb{R}^{L_c \times d}$.

The attention matrix before softmax can be partitioned as:

\begin{equation}
\frac{\mathbf{Q}\mathbf{K}^\top}{\sqrt{d}} = \frac{1}{\sqrt{d}}
\begin{bmatrix}
\mathbf{Q}_s\mathbf{K}_s^\top & \mathbf{Q}_s\mathbf{K}_c^\top \\
\mathbf{Q}_c\mathbf{K}_s^\top & \mathbf{Q}_c\mathbf{K}_c^\top
\end{bmatrix}
\label{eq:attention_blocks}
\end{equation}

Applying the causal mask $\mathbf{M}$ yields:

\begin{equation}
\frac{\mathbf{Q}\mathbf{K}^\top}{\sqrt{d}} + \mathbf{M} = \frac{1}{\sqrt{d}}
\begin{bmatrix}
\mathbf{Q}_s\mathbf{K}_s^\top + \mathbf{M}_s & \mathbf{0} \\
\mathbf{Q}_c\mathbf{K}_s^\top & \mathbf{Q}_c\mathbf{K}_c^\top
\end{bmatrix}
\label{eq:masked_attention}
\end{equation}

where $\mathbf{M}_s$ is the causal mask for the historical subsequence, and the zero block enforces that historical items cannot attend to future candidates.

\subsection{Mathematical Equivalence Proof}

\textbf{Theorem 1.} The decomposed causal self-attention computes the exact same output as the standard causal self-attention.

\begin{proof}
From Equation~\eqref{eq:masked_attention}, the softmax operation can be applied block-wise. Let:

\begin{align}
\mathbf{A}_{ss} &= \operatorname{softmax}\left(\frac{\mathbf{Q}_s\mathbf{K}_s^\top}{\sqrt{d}} + \mathbf{M}_s\right) \\
\mathbf{A}_{cs} &= \operatorname{softmax}\left(\frac{\mathbf{Q}_c\mathbf{K}_s^\top}{\sqrt{d}}\right) \\
\mathbf{A}_{cc} &= \operatorname{softmax}\left(\frac{\mathbf{Q}_c\mathbf{K}_c^\top}{\sqrt{d}}\right)
\end{align}

The full attention output is:

\begin{align}
\operatorname{CausalAttention}(\mathbf{Q}, \mathbf{K}, \mathbf{V}) &=
\begin{bmatrix}
\mathbf{A}_{ss} & \mathbf{0} \\
\mathbf{A}_{cs} & \mathbf{A}_{cc}
\end{bmatrix}
\begin{bmatrix}
\mathbf{V}_s \\
\mathbf{V}_c
\end{bmatrix} \\
&= \begin{bmatrix}
\mathbf{A}_{ss}\mathbf{V}_s \\
\mathbf{A}_{cs}\mathbf{V}_s + \mathbf{A}_{cc}\mathbf{V}_c
\end{bmatrix}
\label{eq:fucc_output}
\end{align}

Our decomposed implementation computes:
\begin{align}
\text{Historical output: } &\mathbf{O}_s = \mathbf{A}_{ss}\mathbf{V}_s \\
\text{Candidate-to-historical: } &\mathbf{C}_{cs} = \mathbf{A}_{cs}\mathbf{V}_s \\
\text{Candidate self-attention: } &\mathbf{C}_{cc} = \mathbf{A}_{cc}\mathbf{V}_c \\
\text{Final candidate output: } &\mathbf{O}_c = \mathbf{C}_{cs} + \mathbf{C}_{cc}
\end{align}

Concatenating $\mathbf{O}_s$ and $\mathbf{O}_c$ yields exactly Equation~\eqref{eq:full_output}. The computational advantage comes from: (1) $\mathbf{A}_{cc}$ is a diagonal matrix due to the causal constraint between candidates, allowing $O(L_c)$ computation; and (2) $\mathbf{A}_{cs}$ has shape $L_c \times L_s$, avoiding the $O(L_c^2)$ computation of full candidate-to-candidate attention.
\end{proof}

%% file: appendix_hp_tables.tex
\section{Detailed Hyper-parameter Analysis Data}
\label{app:hp_tables}

This appendix provides the detailed numerical data corresponding to the hyper-parameter analysis shown in Figure~\ref{fig:hp_merge} in the main text.

\begin{table}[htbp]
\centering
\caption{CA Recall@500 and Latency vs. Sequence Length ($L$)}
\label{tab:seq_length}
\begin{tabular}{@{}ccc@{}}
\toprule
\textbf{Sequence Length ($L$)} & \textbf{Recall@500} & \textbf{NDCG@500} \\
\midrule
64(SA) & 0.1768 & 0.0541 \\
64 & 0.1574 & 0.0459 \\
256 & 0.1670 & 0.0539 \\
512 & 0.2193 & 0.0692 \\
1,000(ours) & \textbf{0.2346} & \textbf{0.0775} \\
\bottomrule
\end{tabular}
\vspace{0.2cm}
\\
\footnotesize{\textit{Note:} The peak CA Recall@500 is achieved at $L=1,000$. Latency increases approximately linearly with sequence length.}
\end{table}

\begin{table}[htbp]
\centering
\caption{CA Recall@500 and QPS vs. Candidate Size ($k_1$)}
\label{tab:candidate_size}
\begin{tabular}{@{}cccc@{}}
\toprule
\textbf{Candidate Size ($k_1$)} & \textbf{Recall@500} & \textbf{NDCG@500}  & \textbf{QPS} \\
\midrule
1,000 & 0.1345 & 0.0398 & 879 \\ 
2,000(ours) & 0.2346 & 0.0775 & 776 \\
5,000 & 0.2419 & 0.0800 & 636 \\
10,000 & 0.2442 & 0.0806 & 450 \\
\bottomrule
\end{tabular}
\vspace{0.2cm}
\\
\footnotesize{\textit{Note:} Diminishing returns begin at $k_1=2,000$, where recall improves only marginally (0.1 percentage points at $k_1=5,000$) while QPS drops significantly (18\% from $k_1=2,000$ to $5,000$).}
\end{table}

\begin{table}[htbp]
\centering
\caption{CA Recall@500 and QPS vs. Top Embedding Dimension ($d_{\text{top}}$)}
\label{tab:embed_dim}
\begin{tabular}{@{}cccc@{}}
\toprule
\textbf{\makecell{Top Embedding \\Dim ($d_{\text{top}}$)}} & \textbf{Recall@500} & \textbf{NDCG@500} & \textbf{QPS} \\
\midrule
32 & 0.2150 & 0.0711   & 1300\\
64 & 0.2299& 0.0758   & 1200 \\
128(ours)& 0.2346 & 0.0775  & 767 \\
256 & 0.2202 &  0.0722  & 500 \\
\bottomrule
\end{tabular}
\vspace{0.2cm}
\\
\footnotesize{\textit{Note:} Recall saturates at $d_{\text{top}}=128$ and declines at 256 due to memory pressure. Using $d_{\text{top}}=64$ trades approximately 3\% recall for a 56\% QPS gain compared to $d_{\text{top}}=128$.}
\end{table}

%% file: appendix_training_details.tex
\section{Training and Implementation Details}
\label{app:training_details}

This appendix provides the complete training, optimization, and deployment configurations for the \textsc{GRank} framework, supplementing the high-level description in Section~\ref{subsubsec:impl_details}. These details are critical for reproducibility and industrial adoption.

\subsection{Training Data \& Sampling Strategy}
We adopt a session-based training paradigm. For each positive user-item pair $(u, i^+)$, we use the user's chronological behavior sequence ending just before $i^+$'s timestamp to construct the model input, preventing any future data leakage.
We employ efficient \textit{in-batch negative sampling}: within each mini-batch of 300 examples, items from all other users serve as negatives for the current user. This yields approximately 299 negatives per positive example at minimal computational cost, facilitating effective contrastive learning.

\subsection{Optimization Configuration}
The model is trained end-to-end using the AdamW optimizer with a learning rate of $1\times10^{-3}$, $\beta_1=0.9$, $\beta_2=0.999$, and a weight decay of $0.01$. Gradient clipping is applied at a global norm of $10.0$.

\subsection{Hardware \& Training Scale}
Training was conducted on a distributed cluster with 2 GPUs, leveraging automatic mixed precision (AMP) to accelerate computation and optimize memory usage. For the largest dataset (Kuaishou), the total training time was approximately 24 hours. 

\subsection{Serving Configuration}
For online inference, the Generator computes a single user embedding per request, while all item tower embeddings are pre-computed offline. The first-stage retrieval is performed via a fast, approximate k-nearest neighbor (KNN) search over these item tower embeddings to obtain a candidate set.
The lightweight Ranker is accelerated using FP16 precision. The complete two-stage pipeline is deployed as a unified service on GPU machines, engineered to meet the industrial P99 latency target of $\leq 100$ms per request.

%% file: appendix_robust.tex
\section{Robustness Analysis Across User and Item Subgroups}

We analyze model robustness under heterogeneous data distributions by evaluating performance across user and item activity subgroups. Users and items are stratified into five groups according to interaction frequency, where \textit{Active\_0} denotes the least active group and \textit{Active\_4} the most active.

\begin{figure}[htbp]
    \centering
    \begin{subfigure}[b]{0.48\textwidth}
        \centering
        \includegraphics[width=\textwidth]{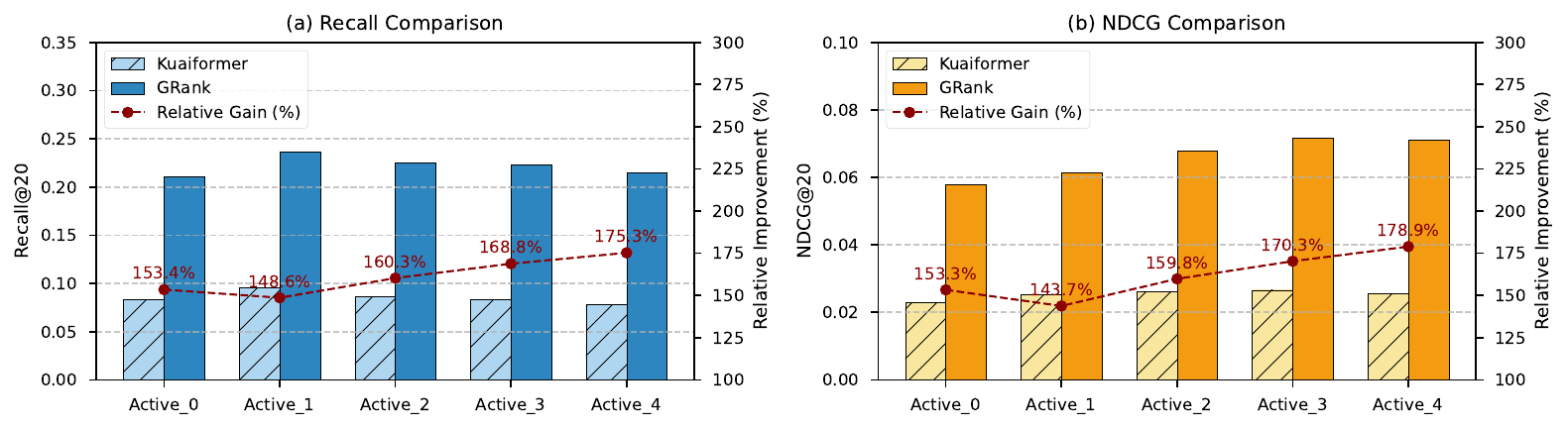}
        \caption{User Activity Groups}
        \label{fig:user_group}
    \end{subfigure}
    \hspace{-0.04\textwidth} %
    \begin{subfigure}[b]{0.48\textwidth}
        \centering
        \includegraphics[width=\textwidth]{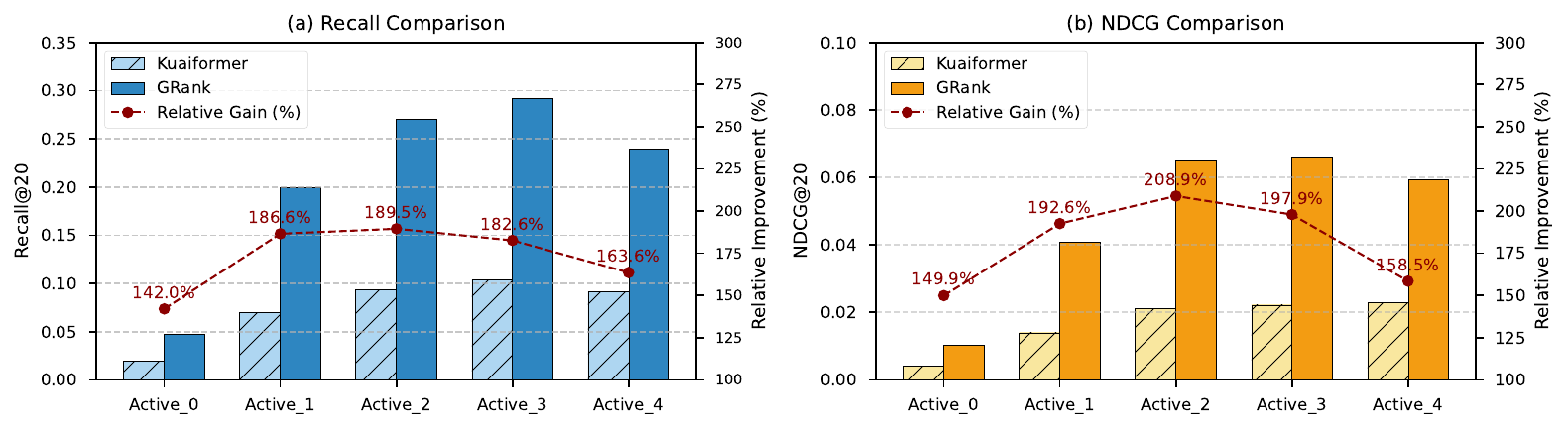}
        \caption{Item Activity Groups}
        \label{fig:item_group}
    \end{subfigure}
    \caption{Robustness evaluation across stratified user and item activity subgroups. Entities are grouped from $Active\_0$ (least active users / long-tail items) to $Active\_4$ (daily active users / popular items). Bars denote absolute metrics (left Y-axis), while red lines represent the relative improvement of GRank over Kuaiformer (right Y-axis). GRank exhibits consistent superiority and strong resilience to interaction sparsity in both user and item dimensions.}
    \label{fig:activity_groups}
\end{figure}

As shown in Table~\ref{fig:user_group}, GRank consistently outperforms Kuaiformer across all user activity groups in terms of both Recall and NDCG. Notably, GRank maintains strong performance for low-activity users, indicating reduced sensitivity to interaction sparsity. In contrast, Kuaiformer exhibits weaker and less stable performance, particularly in sparse regimes.

Similar trends are observed across item activity groups in Table~\ref{fig:user_group}. GRank achieves uniformly higher Recall and NDCG for both low-activity (long-tail) and high-activity items, suggesting improved generalization beyond popular content. Overall, these results demonstrate that GRank is robust to activity-level heterogeneity on both the user and item sides, providing stable recommendation quality under varying data sparsity conditions.

